\documentclass[prd,nofootinbib,superscriptaddress,10pt]{revtex4-2}

\usepackage{amsmath, amssymb, amsthm, graphicx, epsfig, fancyhdr,epsfig, slashed, mathrsfs}

\usepackage{tikzsymbols}
\usepackage{natbib}
\usepackage{float}

\usepackage{mathtools} 

\usepackage{tikz,xcolor}
\usepackage[breaklinks,pdfpagelabels]{hyperref}
\hypersetup{colorlinks=true,linkcolor=blue,urlcolor=blue,citecolor=blue}

\definecolor{lime}{HTML}{A6CE39}
\DeclareRobustCommand{\orcidicon}{
	\begin{tikzpicture}
	\draw[lime, fill=lime] (0,0) 
	circle [radius=0.2] 
	node[white] {{\fontfamily{qag}\selectfont \tiny ID}};
	\draw[white, fill=white] (-0.0625,0.095) 
	circle [radius=0.007];
	\end{tikzpicture}
	\hspace{-2mm}
}

\foreach \x in {A, ..., Z}{\expandafter\xdef\csname orcid\x\endcsname{\noexpand\href{https://orcid.org/\csname orcidauthor\x\endcsname}
			{\noexpand\orcidicon}}
}

\newcommand{\be}{\begin{equation}}
\newcommand{\ee}{\end{equation}}
\newcommand{\bea}{\begin{eqnarray}}
\newcommand{\eea}{\end{eqnarray}}

\newcommand{\bbbone}{\hbox{\rm 1\kern-3pt l}}
\newcommand{\slp}{p\kern-5pt/}

\newcommand{\dAlem}{\hbox{\,\vbox{\hrule height0.3pt\hbox{\vrule width0.3pt
  \hbox{\vbox to8pt{\hbox to8pt{\hfill}\vfill}}\vrule width1.0pt}\hrule
  height1.0pt}\,}}
\newcommand{\dalem}{\hbox{\,\vbox{\hrule height0.2pt\hbox{\vrule width0.2pt
  \hbox{\vbox to5pt{\hbox to5pt{\hfill}\vfill}}\vrule width.6pt%
  \kern-0.1pt}\hrule height0.6pt}\,}}

\def\cob{\color{blue}}
\newcommand{\au}[2]{#1.~#2}
\newcommand{\arX}[1]{\href{http://arxiv.org/abs/#1}{{\cob arXiv:#1}}}

\newcommand{\book}[5]{\emph{#1} (#2, #3, #5)}

\newcommand{\doin}[6]{\href{http://dx.doi.org/#1}{\cob #2\ #3 {\bf #4}, #5 (#6)}}

\newcommand{\doinn}[5]{\href{http://dx.doi.org/#1}{{\cob #2 {\bf #3}, #4 (#5)}}}
\newcommand{\doij}[5]{\href{http://dx.doi.org/#1}{{\cob #2 {\bf #3}, #4 (#5)}}}

\newcommand{\tia}[1]{{#1},}

\begin{document}

\title{Effective potential in non-perturbative gauge theories}

\author{Gianluca Calcagni\orcidD{}}
\email{g.calcagni@csic.es}
\affiliation{Instituto de Estructura de la Materia, CSIC, Serrano 121, 28006 Madrid, Spain}

\author{Marco Frasca\orcidA{}}
\email{marcofrasca@mclink.it}
\affiliation{Rome, Italy}

\author{Anish Ghoshal\orcidB{}}
\email{anish.ghoshal@fuw.edu.pl}

\affiliation{Institute of Theoretical Physics, Faculty of Physics, University of Warsaw, ul. Pasteura 5, 02-093 Warsaw, Poland}

\begin{abstract}
We consider a formalism to describe the false-vacuum decay of a scalar field in gauge theories in non-perturbative regimes. We find that the larger the gauge coupling with respect to the self-coupling of the scalar, the shallower the local minimum of the unstable vacuum, to the point where it disappears. This offers the possibility to obtain a consistent picture of early universe cosmology: at high temperatures, a false-vacuum decay is strongly favoured and the universe naturally evolves towards a stable state.
\end{abstract}

\maketitle

\section{Introduction}

With the advent of gravitational-wave astronomy in the past few years \cite{LIGOScientific:2016lio} and the Higgs boson discovery at LHC and its implication of the vacuum meta-stability \cite{Degrassi:2012ry,Isidori:2001bm} made the possibility to observe a stochastic gravitational-wave background produced from primordial first-order phase transition more concrete than ever (see, e.g, \cite{LISACosmologyWorkingGroup:2022jok} and references therein). Constraints on such a stochastic
gravitational waves background (SGWB) had been given well before the first observations of gravitational waves \cite{2009Natur.460..990A} and presently  are being constantly refined across different frequencies and regions in the parameter space of the models \cite{LISACosmologyWorkingGroup:2022jok}. We are currently looking for more and more accurate calculations of the false-vacuum decay rate in the myriad of beyond-the-Standard-Model theories available, ranging from scenarios described by weak perturbation theory to strongly coupled regimes for Higgs as well as gauge theories. 

In this paper, we make a step further towards our understanding of false-vacuum decay using a technique useful to probe analytically the non-perturbative regime of quantum field theories \cite{Frasca:2015wva,Frasca:2015yva,Chaichian:2018cyv,Frasca:2021yuu,Frasca:2021zyn,Frasca:2022kfy,Frasca:2022lwp,Frasca:2022vvp,Calcagni:2022tls}. After reviewing in Secs.\ \ref{sec01}-\ref{sec03} the non-perturbative techniques we will employ to get our main results, we will consider a theory with scalar and gauge fields\footnote{False vacuum decay rate was discussed in gauge theories in weak perturbation regimes in Refs.\ \cite{Ai:2020sru,Endo:2017gal,Endo:2017tsz,Plascencia:2015pga}.} (Sec.\ \ref{sec2}) and compute the effective potential and the fluctuations for the scalar (Sec.\ \ref{sec3}), finding that the height of the minima depends on the interplay between the scalar self-coupling $\lambda$ and the gauge coupling $g$. Although we will work in Minkowski spacetime without gravity, these results can pave the way to a natural picture of the evolution of the early universe from an unstable to a stable vacuum state (Sec.\ \ref{sec4}).

\section{Dyson-Schwinger equations and Bender-Milton-Savage technique}
\label{sec01}

In the following, we present the Bender-Milton-Savage technique \cite{Bender:1999ek}. This permits one to obtain the full hierarchy of Dyson-Schwinger equations in a partial differential equation (PDE) form. Four our aims, it is more convenient to work in the Euclidean metric.

Let us consider the partition function for a scalar field
\begin{equation}
    Z[j]=\int[D\phi]e^{-S(\phi)+\int d^4xj(x)\phi(x)}.
\end{equation}
For the one-point function, we get
\be
\left\langle\frac{\delta S}{\delta\phi(x)}\right\rangle=j(x)\,,
\ee
where
\be
\left\langle\ldots\right\rangle=\frac{\int[D\phi]\ldots e^{-S(\phi)+\int d^4xj(x)\phi(x)}}{\int[D\phi]e^{-S(\phi)+\int d^4xj(x)\phi(x)}}.
\ee
After that, we can complete the procedure by setting $j=0$. The next step is to derive the above equation for the $j$-dependent one-point function to obtain the equation for the two-point function. Since the definition of the $n$-point function is
\be
\langle\phi(x_1)\phi(x_2)\ldots\phi(x_n)\rangle=\frac{\delta^n\ln(Z[j])}{\delta j(x_1)\delta j(x_2)\ldots\delta j(x_n)}\,,
\ee
one has
\be
\frac{\delta G_k(\ldots)}{\delta j(x)}=G_{k+1}(\ldots,x).
\ee

As an example, take a $\phi^4$ theory, for which one has
\be
S=\int d^4x\left[\frac{1}{2}(\partial\phi)^2-\frac{\lambda}{4}\phi^4\right],
\ee
so that
\be
\label{eq:G_1}
\partial^2\langle\phi\rangle+\lambda\langle\phi^3(x)\rangle = j(x).
\ee
The following equation just holds
\be
Z[j]\partial^2G_1^{(j)}(x)+\lambda\langle\phi^3(x)\rangle = j(x).
\ee
By the definition of the one-point function, one gets
\be
Z[j]G_1^{(j)}(x)=\langle\phi(x)\rangle.
\ee
Now we derive with respect to $j(x)$ to obtain
\be
Z[j][G_1^{(j)}(x)]^2+Z[j]G_2^{(j)}(x,x)=\langle\phi^2(x)\rangle,
\ee
and after another derivation step we have
\be
Z[j][G_1^{(j)}(x)]^3+3Z[j]G_1^{(j)}(x)G_2(x,x)+Z[j]G_3^{(j)}(x,x,x)=\langle\phi^3(x)\rangle.
\ee
Inserting it into Eq.\ (\ref{eq:G_1}) yields
\be
\label{eq:G1_j}
\partial^2G_1^{(j)}(x)+\lambda[G_1^{(j)}(x)]^3+3\lambda G_2^{(j)}(0)G_1^{(j)}(x)+G_3^{(j)}(0,0)=Z^{-1}[j]j(x)\,.
\ee
We realize that, by the effect of renormalization, a mass term appeared. We uncover here a term due to mass renormalization. Therefore, setting $j=0$, one gets the first Dyson-Schwinger equation into differential form:
\be
\label{eq:g10}
\partial^2G_1(x)+\lambda[G_1(x)]^3+3\lambda G_2(0)G_1(x)+G_3(0,0)=0.
\ee

By deriving Eq.\ (\ref{eq:G1_j}) again with respect to $j(y)$, we get
\be
\begin{split}
&\partial^2G_2^{(j)}(x,y)+3\lambda[G_1^{(j)}(x)]^2G_2^{(j)}(x,y)+
\nonumber \\
&3\lambda G_3^{(j)}(x,x,y)G_1^{(j)}(x)
+3\lambda G_2^{(j)}(x,x)G_2^{(j)}(x,y)
+G_4^{(j)}(x,x,x,y)=\nonumber \\
&Z^{-1}[j]\delta^4(x-y)+j(x)\frac{\delta}{\delta j(y)}(Z^{-1}[j]).
    \end{split}
\ee
Inserting $j=0$, the equation for the two-point function takes the form
\be
\partial^2G_2(x,y)+3\lambda[G_1(x)]^2G_2(x,y)+
3\lambda G_3(0,y)G_1(x)
+3\lambda G_2(0)G_2(x,y)
+G_4(0,0,y)=
\delta^4(x-y).
\ee
This procedure can be iterated to any desired order providing all the hierarchy of Dyson-Schwinger equations in PDE form.


In order to understand our approach in a properly framed setting, we consider the general relation between cumulants and averages that characterizes the evaluation we accomplish of the partition function of the theory. We can generate any correlation function by the generating functional that is the equivalent to the generating function of the relation between cumulants and averages in probability theory
\be
\langle e^{\int d^4xj\phi}\rangle Z^{-1}[j]=\exp\left[\sum_{n=1}^\infty\frac{1}{n!}G_n^j(x_1,x_2,\ldots,x_n)j(x_1)j(x_2)\ldots,j(x_n)\right].
\ee
Our solution corresponds to the Gaussian limit that can always be found and so, higher order correlation functions for $n>2$ will be given by products of $G_1$ and $G_2$. As the 1P-correlation function decays at large coupling, higher products of them are more and more damped making a very good approximation to truncate to the very first contributions in the partition function. We will use this result in our evaluation of the effective potential.
%

\section{Dyson-Schwinger equations for one-point and two-point functions of gauge theory}
\label{sec02}

Consider the Lagrangian, with the convention that $\mu,\nu,\lambda,\ldots$ are the relativistic indexes and $a,b,c,\ldots$ are the group indices,
\be\label{teor1}
{\cal L}_{YM}=-\frac{1}{4}F_{\mu\nu}^aF^{a\mu\nu}+L_{\rm GF}+L_{\rm FP},
\ee
where
\begin{equation}
L_{\rm GF} = 
\frac{1}{2 \xi}A_{\mu}^a\partial^{\mu}\partial^{\nu}A_{\nu}^{a}, 
\label{NLgauge}
\end{equation}
being $\xi$ an arbitrary constant, dubbed the gauge fixing parameter, to be properly chosen, and
\be
L_{\rm FP}=-\bar{c}^ae^{-f(\Box)}(\partial^\mu D_{\mu}^{ab})c^b,
\label{NLghost}
\ee
where $c,\ \bar{c}$ are the ghost fields and $D_{\mu}$ the covariant derivative. We will obtain the Dyson-Schwinger equations for the Yang-Mills field for the one-point and two-point functions using the Bender-Milton-Savage technique described in the preceding section.

For the one-point functions, after averaging the equations of motion, we get
\begin{eqnarray}
\Box G_{1\mu}^{(j)a}+gf^{abc}\left\langle\partial_\nu\left[{\bar A}^{b}_\mu
A^{c\nu}\right]\right\rangle+&& \nonumber \\
gf^{abc}\left\langle\left[ A^{b\nu}
(\partial_\mu A^{c}_\nu-\partial_\nu A^{c}_\mu)\right]\right\rangle&& \nonumber \\
g^2f^{abc}f^{cde}
\left\langle\left[A^{b\nu}A^{d}_\nu
A^{e}_\mu\right]\right\rangle+&& \nonumber \\
+gf^{abc}\left\langle \bar{c}^b\partial_\mu c^c\right\rangle&=&j^a_\mu,
\end{eqnarray}
and for the ghost
\be
-\Box P_1^{(\eta)a} +gf^{abc}\left\langle\left(A_\mu^c\right)\partial^\mu c^b\right\rangle=\eta^a,
\ee
where we used the definitions
\bea
\label{eq:defs}
G_{1\mu}^{(j)a}(x)&=&Z^{-1}\langle A_\mu^a(x)\rangle \nonumber \\
P_1^{(\eta)a}(x)=&=&Z^{-1}\langle c^a(x)\rangle.
\eea
The same should hold for ${\bar c}^a$ yielding ${\bar P}_1^{(\eta)a}(x)$. 
In order to evaluate the averages, we consider the above definitions rewritten as
\bea
Z[j,\eta,{\bar\eta}]G_{1\mu}^{(j)a}(x)&=&\langle A_\mu^a(x)\rangle \nonumber \\
Z[j,\eta,{\bar\eta}]P_1^{(\eta)a}(x)&=&\langle c^a(x)\rangle.
\eea
The superscripts $(j)$ and $(\eta)$ are there to remind us of the explicit dependence on the currents. Let us derive once with respect to $j(x)$ on the first equation to get
\be
\label{eq:f1}
ZG_{2\mu\nu}^{(j)ab}(x,x)+
ZG_{1\mu}^{(j)a}(x)G_{1\nu}^{(j)b}(x)=
\langle A_\mu^a(x)A_\nu^b(x)\rangle.
\ee
Applying the spacetime derivative $\partial^\nu$, we obtain
\be
Z\partial^\nu G_{2\mu\nu}^{(j)ab}(x,x)+
Z\partial^\nu G_{1\mu}^{(j)a}(x)G_{1\nu}^{(j)b}(x)=
\langle \partial^\nu A_\mu^a(x)A_\nu^b(x)\rangle.
\ee
We further derive Eq.\ (\ref{eq:f1}) with respect to $j^{c\nu}$ to get
\bea
ZG_{2\mu\nu}^{(j)ab}(x,x)G_1^{(j)\nu c}(x)
+ZG_{3\mu\nu}^{(j)abc\nu}(x,x,x)+
\nonumber \\
ZG_{1\mu}^{(j)a}(x)G_{1\nu}^{(j)b}(x)G_1^{(j)\nu c}(x)+ZG_{2\mu}^{(j)ac\nu}(x)G_{1\nu}^{(j)b}+\nonumber \\
ZG_{2\nu}^{(j)bc\nu}(x)G_{1\mu}^{(j)a}(x)=
\langle A_\mu^a(x)A_\nu^b(x)A^{c\nu}(x)\rangle,
\eea
and we need to do the same for the ghost field. From Eq.\ (\ref{eq:defs}), we write
\be
\label{eq:P1}
Z[j,\eta,\bar\eta]P_1^{(\eta)a}(x)=\langle c^a(x)\rangle.
\ee
After deriving with respect to $\partial_\mu$ and then with respect to $\bar\eta$, one has
\be
Z{\bar P}_1^{(\eta)b}(x)\partial^\mu P_1^{(\eta)a}(x)+Z\partial^\mu K_2^{(\eta)ab}(x,x)=
\langle {\bar c}^b\partial^\mu c^a(x)\rangle,
\ee
where we have naturally defined the two-point function 
\be
K_2^{(\eta)ab}(x,y)=\frac{1}{Z}\frac{\delta P_1^{(\eta)a}(x)}{\delta \eta^b(y)},
\ee
with the other two-point function being
\be
J_{2\mu}^{(\eta,j)ab}(x,y)=\frac{1}{Z}\frac{\delta P_1^{(\eta)a}(x)}{\delta j^{b\mu}(y)}.
\ee
Differentiating Eq.\ (\ref{eq:P1}) with respect to $j^{b\mu}(x)$, the result is
\be
Ze^{\frac{1}{2}f(\Box)}G_{1\mu}^{(j)b}(x)\partial^\mu P_1^{(\eta)a}(x)+Z\partial^\mu J_{2\mu}^{(\eta,j)ab}(x,x)=
\langle A^b_\mu(x)\partial^\mu c^a(x)\rangle.
\ee
Collecting everything together, one has
\begin{eqnarray}
\label{eq:G1j}
\Box G_{1\mu}^{(j)a}+gf^{abc}
\partial^\nu\left[
G_{2\mu\nu}^{(j)bc}(x,x)+
G_{1\mu}^{(j)b}(x)G_{1\nu}^{(j)c}(x)
\right]-&& \nonumber \\
gf^{abc}
\left[ 
\partial^\nu G_{2\mu\nu}^{(j)bc}(x,x)+
\partial^\nu G_{1\mu}^{(j)b}(x)G_{1\nu}^{(j)c}(x)
\right]-
&& \nonumber \\
gf^{abc}
\left[ 
\partial_\mu G_{2\nu}^{(j)bc\nu}(x,x)+
\partial_\mu G_{1\nu}^{(j)b}(x)e^{\frac{1}{2}f(\Box)}G_{1}^{(j)c\nu}(x)
\right]+
&& \nonumber \\
g^2f^{abc}f^{cde}
\left[
G_{2\mu\nu}^{(j)bd}(x,x)G_1^{(j)\nu e}(x)
+\partial^\nu G_{3\mu\nu}^{(j)bde\nu}(x,x,x)+
\right.
\nonumber \\
G_{1\mu}^{(j)b}(x)G_{1\nu}^{(j)d}(x)G_1^{(j)\nu e}(x)+G_{2\mu}^{(j)be\nu}(x,x)G_{1\nu}^{(j)d}(x)+\nonumber \\
\left.
G_{2\nu}^{(j)de\nu}(x,x)G_{1\mu}^{(j)b}(x)
\right]-&& \nonumber \\
gf^{abc}
\left\{
{\bar P}_1^{(\eta)b}(x)\left[\partial_\mu P_1^{(\eta)c}(x)\right]+\partial_\mu\left[K_2^{(\eta)bc}(x,x)\right]
\right\}
&=& j^a_\mu\,.
\end{eqnarray}
The equations of the local theory given in \cite{Frasca:2015yva} are easily obtained by setting the non-locality factor to 1, corresponding to the local limit $M \rightarrow \infty$
For the ghost field, it is
\be
-\Box P_1^{(\eta)c} 
-gf^{abc}G_{1\mu}^{(j)a}(x)\partial^\mu P_1^{(\eta)b}(x)-gf^{abc}\partial^\mu J_{2\mu}^{(\eta,j)ab}(x,x)
=\eta^c.
\ee
After setting all the currents to zero, the Dyson-Schwinger equations for the one-point functions are obtained in the form given in the text.

Differentiating Eq.\ (\ref{eq:G1j}) with respect to $j^{\lambda h}(y)$, one obtains
\begin{eqnarray}
\label{eq:G2j}
\Box G_{2\mu\lambda}^{(j)ah}(x,y)+gf^{abc}
\partial^\nu\left[
G_{3\mu\nu\lambda}^{(j)bch}(x,x,y)+
G_{2\mu\lambda}^{(j)bh}(x,y)\times
\right.&&
\nonumber \\
\left.
G_{1\nu}^{(j)c}(x)+
+G_{1\mu}^{(j)b}(x)
G_{2\nu\lambda}^{(j)ch}(x)
\right]-&& \nonumber \\
gf^{abc}
\left[ 
\partial^\nu G_{2\mu\nu\lambda}^{(j)bch}(x,x,y)+
\partial^\nu G_{2\mu\lambda}^{(j)bh}(x,y)G_{1\nu}^{(j)c}(x)+
\right.
&& \nonumber \\
\left.
\partial^\nu G_{1\mu}^{(j)b}(x)G_{2\nu\lambda}^{(j)ch}(x,y)
\right]-
&& \nonumber \\
gf^{abc}
\left[
\partial_\mu G_{3\nu\lambda}^{(j)bch\nu}(x,x,y)+
\partial_\mu G_{2\nu\lambda}^{(j)bh}(x,y)G_{1}^{(j)c\nu}(x)+
\right.&&
\nonumber \\
\left.
\partial_\mu G_{1\nu}^{(j)b}(x)G_{2\lambda}^{(j)ch\nu}(x,y)
\right]+
&& \nonumber \\
g^2f^{abc}f^{cde}
\left[
G_{3\mu\nu\lambda}^{(j)bdh}(x,x,y)G_1^{(j)\nu e}(x)+
\right.&&
\nonumber \\
G_{2\mu\nu}^{(j)bd}(x,x)G_{2\lambda}^{(j)\nu eh}(x,y)
+\partial^\nu G_{4\mu\nu\lambda}^{(j)bdeh\nu}(x,x,x,y)+
\nonumber \\
G_{2\mu\lambda}^{(j)bh}(x,y)G_{1\nu}^{(j)d}(x)e^{\frac{1}{2}f(\Box)}G_1^{(j)\nu
e}(x)+&&
\nonumber \\
G_{1\mu}^{(j)b}(x)G_{2\nu\lambda}^{(j)dh}(x,y)G_1^{(j)\nu
e}(x)+&&
\nonumber \\
G_{1\mu}^{(j)b}(x)G_{1\nu}^{(j)d}(x)G_{2\lambda}^{(j)\nu
eh}(x,y)+&&
\nonumber \\
G_{3\mu\lambda}^{(j)beh\nu}(x,x,y)G_{1\nu}^{(j)d}(x)+&& 
\nonumber \\
G_{2\mu}^{(j)be\nu}(x,x)G_{2\nu\lambda}^{(j)dh}(x,y)+&&
\nonumber \\
\left.
G_{3\nu\lambda}^{(j)deh\nu}(x,x,y)G_{1\mu}^{(j)b}(x)
+G_{2\nu}^{(j)de\nu}(x,x)G_{2\mu\lambda}^{(j)bh}(x,y)
\right]-&& \nonumber \\
gf^{abc}
\left\{
{\bar J}_{2\lambda}^{(\eta,j)bh}(x,y)\left[\partial_\mu P_1^{(\eta)c}(x)\right]\right.+&& \nonumber \\
\left.{\bar P}_1^{(\eta)b}(x)\left[\partial_\mu J_{2\lambda}^{(\eta)ch}(x,y)\right]
+\partial_\mu\left[W_{3\lambda}^{(\eta,j)bch}(x,x,y)\right]
\right\}=&& 
\delta^{ah}\eta_{\mu\lambda}\delta^4(x-y),
\end{eqnarray}
where the three-point function $W_3$ is
\be
W_{3\lambda}^{(\eta,j)abc}(x,y,z)=Z^{-1}\frac{\delta K_2^{(\eta)ab}(x,y)}{\delta j^{\lambda c}(z)}.
\ee
Similarly, starting from the one-point function for the ghost and deriving it with respect to $\eta^{h}(y)$, we get
\bea
&-\Box K_2^{(\eta)ch}(x,y)
-igL_{2\mu}^{(\eta,j)ah}(x,y)\partial^\mu P_1^{(\eta)b}(x)\nonumber \\
&-igf^{abc}G_{1\mu}^{(j)a}(x)\partial^\mu K_2^{(\eta)bh}(x,y)
-igf^{abc}\partial^\mu W_{3\mu}^{(\eta,j)abh}(x,x,y) \nonumber \\
&=\delta^{ch}\delta^4(x-y),
\eea
where
\be
L_{2\mu}^{(\eta,j)ab}(x,y)=\frac{\delta G_1^{(j)a}(x)}{\delta \eta^b(y)}.
\ee
Deriving with respect to $j^{h\nu}(y)$, one has the equation for $J_2$ in the form
\bea
&-\Box J_2^{(\eta)ch\nu}(x,y) 
-igf^{abc}G_{2\mu\nu}^{(j)ah}(x,y)\partial^\mu P_1^{(\eta)b}(x) \nonumber \\
&-igf^{abc}G_{1\mu}^{(j)a}(x)\partial^\mu J_2^{(\eta,j)bh\nu}(x,y) \nonumber \\
&-igf^{abc}\partial^\mu J_{3\mu}^{(\eta,j)abh}(x,x,y)=0,
\eea
with the three-point function
\be
J_{3\mu}^{(\eta,j)abc}(x,y,z)=\frac{\delta J_{2\mu}^{(\eta,j)ab}(x,y)}{\delta j^{c\mu}(z)}.
\ee
We can recover the equations to be solved by setting all the currents to zero. For the one-point correlation function, this yields
\begin{eqnarray}\label{EL2}
\lefteqn{\dAlem G_{1\mu}^{(j)a}(x)+j_\mu^a(x)
  \ =}\nonumber\\
  &&-gf_{abc}\Big\{\partial^\nu
  \Big(G_{2\mu\nu}^{(j)bc}(x,x)+G_{1\mu}^{(j)b}(x)G_{1\nu}^{(j)c}(x)\Big)
  +(\partial_\mu G_{2\nu b}^{(j)c\nu}(x,x)-\partial_\nu G_{2\mu b}^{(j)c\nu}(x,x))
  \strut\nonumber\\&&\strut\qquad\qquad
  +G_{1b}^{(j)\nu}(x)
  (\partial_\mu G_{1\nu}^{(j)c}(x)-\partial_\nu G_{1\mu}^{(j)c}(x))\Big\}
  \strut\nonumber\\&&\strut-g^2f_{abc}f_{cde}\Big\{
  \Big( G_{3\mu\nu b}^{(j)de\nu}(x,x,x)+G_{2\mu b}^{(j)d\nu}(x,x)G_{1\nu}^{(j)e}(x)+G_{1\mu}^{(j)d}(x)G_{2\nu b}^{(j)e\nu}(x,x)\Big)
  \strut\nonumber\\&&\strut\qquad\qquad
  +G_{1b}^{(j)\nu}(x)
  \Big(G_{2\mu\nu}^{(j)de}(x,x)+G_{1\mu}^{(j)d}(x)G_{1\nu}^{(j)e}(x)\Big)\Big\}.
\end{eqnarray}
The equation of motion for the two-point function is obtained by varying
with respect to $j^\lambda_h(y)$:
\begin{eqnarray}
\lefteqn{\dAlem G_{2\mu\lambda}^{(j)ah}(x,y)-\delta^{ah}
  \eta_{\mu\lambda}\delta^{(4)}(x-y)\ =}\nonumber\\
  &&-gf_{abc}\Big\{\partial^\nu
  \Big(G_{3\mu\nu\lambda}^{(j)bch}(x,x,y)
  +G_{2\mu\lambda}^{(j)bh}(x,y)G_{1\nu}^{(j)c}(x)
  \strut\nonumber\\&&\strut\qquad\qquad
  +G_{1\mu}^{(j)b}(x)G_{2\nu\lambda}^{(j)ch}(x,y)\Big)
  \strut\nonumber\\&&\strut\qquad
  +(\partial_\mu G_{3\nu b\lambda}^{(j)c\nu h}(x,x,y)-\partial_\nu G_{3\mu b\lambda}^{(j)c\nu h}(x,x,y))\strut\nonumber\\&&\strut\qquad
  +\Big(G_{2b\lambda}^{(j)\nu h}(x,y)(\partial_\mu G_{1\nu}^{(j)c}(x)
  -\partial_\nu G_{1\mu}^{(j)c}(x))
  \strut\nonumber\\&&\strut\qquad\qquad
  +G_{1b}^{(j)\nu}(x)(\partial_\mu G_{2\nu\lambda}^{(j)ch}(x,y)-\partial_\nu G_{2\mu\lambda}^{(j)ch}
  (x,y))]\Big)\Big\}\strut\nonumber\\&&\strut
  -g^2f_{abc}f_{cde}\Big\{\Big(G_{4\mu\nu b\lambda}^{(j)de\nu h}(x,x,x,y)]
  \strut\nonumber\\&&\strut\qquad\qquad
  +G_{3\mu\nu b}^{(j)d\nu h}(x,x,y)G_{1\nu}^{(j)e}(x)
  +G_{2\mu b}^{(j)d\nu}(x,x)G_{2\nu\lambda}^{(j)eh}(x,y)
  \strut\kern-25pt\nonumber\\&&\strut\qquad\qquad
  +G_{2\mu\lambda}^{(j)dh}(x,y)G_{2\nu b}^{(j)e\nu}(x,x)]
  +G_{1\mu}^{(j)d}(x)G_{3\nu b\lambda}^{(j)e\nu h}(x,x,y)\Big)
  \strut\kern-32pt\nonumber\\&&\strut\qquad
  +G_{2b\lambda}^{(j)\nu h}(x,y)
  (G_{2\mu\nu}^{(j)de}(x,x)+G_{1\mu}^{(j)d}(x)G_{1\nu}^{(j)e}(x))
  \strut\nonumber\\&&\strut\qquad
  +G_{1b}^{(j)\nu}(x)\Big(G_{3\mu\nu\lambda}^{(j)deh}(x,x,y)\strut\\&&\strut\qquad\qquad
  +G_{2\mu\lambda}^{(j)dh}(x,y)G_{1\nu}^{(j)e}(x)
  +G_{1\mu}^{(j)d}(x)G_{2\nu\lambda}^{(j)eh}(x,y)\Big)\Big\}.
  \nonumber
\end{eqnarray}

\section{Choice of the one-point correlation functions}\label{sec03}

Our chosen solution for Eq.~(\ref{eq:g10}), that is our representation of the vacuum of the theory, can be written as the so-called Fubini-Lipatov instanton \cite{Fubini:1976jm,Lipatov:1976ny}
\begin{equation}
\label{solG1}
    G_{1\phi}(x)=\sqrt{\frac{2\mu^4}{m_\phi^2+\sqrt{m_\phi^4+2\lambda\mu^4}}}{\rm sn}\left(p\cdot x+\chi,\frac{-m_\phi^2+\sqrt{m_\phi^4+2\lambda\mu^4}}{-m_\phi^2-\sqrt{m_\phi^4+2\lambda\mu^4}}\right) \,,
\end{equation}
where $\mu$ and $\chi$ are arbitrary integration constants (dimension-one energy scale and dimensionless phase, respectively), sn$(\xi,\nu^2)$ is a Jacobi elliptic function and, provided we define $m_\phi^2 \equiv 3\lambda G_2(0)$, it must be $G_3(x,x,x)=0$, that is shown to be consistent by inspection {\sl a posteriori} using our solution and the two-point correlation function (see below). The four-momentum $p$ of the quasi-particle satisfies 
\begin{equation}
\label{eq:disp}
    p^2=m_\phi^2+\frac{\lambda\mu^4}{m_\phi^2+\sqrt{m_\phi^4+2\lambda\mu^4}} \,.
\end{equation}
It must be emphasized that the momentum $p$ entering into the dispersion relation arises by the integration of the equation for $G_1$ and does not correspond to the momentum of asymptotic states in perturbative quantum field theory. Once again, we point out that this choice is neither unique nor are we able to prove that it is the best one. We choose it because it is in a closed analytical form and all other results we will obtain from it will have the same property. This will give us some interesting understanding of non-perturbative physics.

This choice extends naturally to the Yang-Mills case using the so-called ``mapping theorem.'' We consider a particular class of solutions of the one-point correlation function equation that are mapped onto the scalar field solution as
\be
\label{eq:solG_YM}
 G_{1\mu}^{a}(x)=\eta^a_\mu G_{1\phi}(x),
\ee
where $\eta^a_\mu$ is the polarization vector with
$\eta_\mu^a\eta^\mu_b=\delta_{ab}$. This theorem was proven in \cite{Frasca:2009yp} and is nowadays accepted in the mathematicians' community.\footnote{Terence Tao (Fields medalist) initially questioned this theorem and one of us (M.F.) was asked to publish a correction on a refereed journal. This was done in \cite{Frasca:2009yp}. Tao's acceptance was published in \url{https://dispersivewiki.org/DispersiveWiki/index.php?title=Talk:Yang-Mills_equations} after it was made clear that the mapping is just an asymptotic one, as also stated in the main text.} It is an identity only for our gauge choice, otherwise it can be seen as an asymptotic mapping that holds in the limit of a strongly coupled theory.

We emphasize that these choices represent our specific understanding of the vacuum of these two theories and that different choices are also possible and could be proven valid as well. The specificity of the \emph{Ansatz} is sometimes perceived as an issue, or even a flaw, by researchers familiar with perturbative field theory.
However, we stress that one cannot extend such back-of-the-envelope considerations to nonperturbative field theory and that general or even just useful conclusions about complex aspects of the theory do not have to, or even cannot, be based on too general \emph{Ans\"atze} for vacuum solutions. In this respect, the instantonic solution \eqref{solG1} does not have any handicap \emph{a priori} with respect to others.
The fact that it breaks translation invariance should be treated under the same stance, for two reasons. First, Poincar\'e-invariant vacua are a widely used postulate in perturbative field theory but there is no physical reason why one should not consider alternatives in a nonperturbative setting. Second, we expect that no experiment could be able to detect this breaking of Poincar\'e symmetry coming from the one-point function, not only because $G_1$ is not a physical observable, but also because the two-point correlation function, the one entering into the physical processes of these theories through the LSZ theorem, does not break Poincar\'e invariance at all, as proven in \cite{Frasca:2015yva}. Besides, both these theories, for this choice of the vacuum solution, display confinement, thus hiding possible symmetry violations at very small energy scales, if any \cite{Chaichian:2018cyv,Frasca:2023qii}.

Using the mapping theorem, the equation for one-point and two-point correlation functions of the Yang-Mills theory take the simplified form
\begin{equation}
\label{eq:g101}
   \partial^2 G_1(x)+Ng^2\left([G_1(x)]^3+3G_2(0)G_1(x)+G_3(x,x,x)\right)=0 \,,
\end{equation}
and
\begin{eqnarray}
\label{eq:g201}
   &&\partial^2G_2(x-y)+Ng^2\left(3[G_1(x)]^2G_2(x-y)+3G_2(0)G_2(x-y)\right. \nonumber \\
	&&\left.+3G_3(x,x,y)G_1(x)+G_4(x,x,x,y)\right)=\delta^4(x-y) \,,
\end{eqnarray}
for an $SU(N)$ theory. 
It is important to note a relevant property of non-Abelian gauge theories that was obtained by lattice computations \cite{Bogolubsky:2007ud,Cucchieri:2007md,Oliveira:2007px} and that we recover in our exact solution: The ghost sector decouples from the theory and behaves like that of a free particle. Such a behavior has been dubbed ``decoupling solution'' for the propagators of the theory in literature. This will be very helpful in our analysis to simplify computations.

\section{Model and solutions}\label{sec2}


We will now apply the formalism described in the preceding sections, for our exact solution, to a scalar field coupled to an $SU(N)$ gauge field. In our analysis, the relevance of the choice of the vacuum solution is that, in principle, we are able to compute all the $n$-point correlation functions in analytical closed form. This means that we know the partition function when given in the form \cite{EWeinb}
\begin{equation}
    Z[j] = \sum_n \frac{1}{n!}\int d^4x_1 \cdots d^4x_n \,G_{n}(x_1, \dots,x_n)
       j(x_1) \cdots j(x_n)
\label{WofJ}
\end{equation}
that generalizes immediately to the Yang-Mills theory. Therefore, we assume that a quantum field theory is exactly solved when one is able to compute all the correlation functions $G_n$ in closed analytical form. In our case, this task is accomplished by solving the set of Dyson-Schwinger set of PDEs we obtained using the Bender-Milton-Savage technique.

\subsection{Lagrangian and method}

We consider the following Lagrangian:
\be\label{teor}
{\cal L}=-\frac{1}{4}F_{\mu\nu}^aF^{a\mu\nu}+\frac{1}{2}(\partial_\mu\phi-igT^aA_\mu^a\phi)^2-V(\phi)+L_{\rm GF}+L_{\rm FP},
\ee
where $F_{\mu\nu}^a=\partial_\mu A_\nu^a-\partial_\nu A^a_\nu+gf^{abc}A^b_\mu A^c_\nu$, provided that $[T^a,T^b]=if^{abc}T^c$ with $f^{abc}$ the structure constants of the gauge group. Here, Greek indices indicate spacetime directions and Latin indices run on the internal gauge group. $V(\phi)$ is the potential of the scalar field with a false vacuum. $L_{\rm GF}$ is the gauge fixing part and $L_{\rm FP}$ is the contribution of Faddeev--Popov ghosts $c$ and $\bar c$. We use $(+,-,-,-)$ signature.

In this paper, we will evaluate the partition function of the scalar field coupled to a non-Abelian gauge field. The partition function is a functional of the currents $j_\phi$ and $j_A^{a\mu}$ and $\eta^a,\ {\bar\eta}^a$. The corresponding correlation functions of the theory are defined through the functional Taylor series, that always exists as the currents are arbitrary functions introduced just for this aim, and are obtained as the kernels at the different orders of the functional Taylor series.

\subsection{Solutions}

Let us consider the partition function
\be
Z[j_A,j_\phi,\bar\eta,\eta]=
\int[dA][d\phi][d\bar{c}][dc]e^{i\int d^4xL+i\int d^4x[j_A^{a\mu}A^a_\mu+j_\phi\phi+\bar\eta^ac^a+\bar{c}^a\eta^a]}.
\ee
Then, we notice that
\be
(D\phi)^2=(\partial\phi)^2-2igA^a_\mu(\phi T^a\partial^\mu\phi)-g^2A_\mu^aA^{b\mu}(T^a\phi)(T^b\phi),
\ee
where repeated gauge indices are contracted. For the gauge sector, we observe that the ghost sector decouples for the correlation functions and we can consider the expansion
\be 
A_\mu^a[x;j]=\eta_\mu^a G_1(x)+\int d^4y G_{2\mu\nu}^{ab}(x-y)j^{b\nu}(y)+O(j^2),
\ee
where $\eta_\mu^a$ is a constant tensor with one spacetime and one gauge index and we sum over the gauge index $b$. Then we can write the partition function as
\bea
Z[j_\phi]&=&Z_0
\int[d\phi]e^{i\int d^4x\left[
\frac{1}{2}(\partial\phi)^2-V(\phi)-ig\left(-i\frac{\delta}{\delta j_\mu^a(x)}\right)(\phi T^a\partial\phi)-\frac{1}{2}g^2\left(-i\frac{\delta}{\delta j_\mu^a(x)}\right)\left(-i\frac{\delta}{\delta j^{b\mu}(x)}\right)(T^a\phi)(T^b\phi)+j_\phi\phi\right]}\nonumber\\
&& \times e^{i\int d^4x\eta_\mu^a G_1(x)j^{a\mu}(x)+\frac{i}{2}\int d^4xd^4yj_\mu^a(x)G_2^{ab\mu\nu}(x-y)j_\nu^b(y)}+O(j^3).
\eea
From this partition function, we can get an effective scalar theory given by
\be
Z_\phi[j_\phi]=Z_0
\int[d\phi]e^{i\int d^4x\left[
\frac{1}{2}(\partial\phi)^2-V(\phi)-igG_1(x)(\phi\eta_\mu^aT^a\partial^\mu\phi)
+\frac{1}{2}g^2\eta_\mu^a\eta^{b\mu}[G_1(x)]^2(\phi T^a\partial\phi)
(\phi T^b\partial\phi)
+\frac{1}{2}g^2G_{2\mu}^{ab\mu}(0)(T^a\phi)(T^b\phi)\right]}.
\ee
We see that the presence of the interaction with the gauge field yields a mass term to the scalar field in a first approximation for a strongly coupled theory. Such an approximation permits us to discard the term with $G_1$ that goes like $(Ng^2)^{-\frac{1}{4}}$. We notice that this result is gauge-independent as the propagator of the gauge field enters as $G_{2\mu}^{ab\mu}(0)$ evaluated at zero momentum. This implies that any projector due to the gauge choice entering into the definition does not contribute and we have no dependence on it.

%
Our one-point solution will be given by Eq.\ (\ref{eq:solG_YM}) combined with Eq.\ (\ref{solG1}). For the two-point correlation function, we do not repeat the derivation here (e.g., see \cite{Frasca:2015yva,Chatterjee:2023ehr}). 
In the Landau gauge, the two-point function can be written as
\be
G_{2\mu}^{ab\mu}(x-y)=\delta_{ab}\left(g_{\mu\nu}-\frac{\partial_\mu\partial_\nu}{\partial^2}\right)\Delta(x-y),
\ee
and the function $\Delta(x-y)$ is given in momentum space, for the Euclidean metric, as
\begin{equation}\label{Delta0mom}
\Delta(p)=\sum_{n=0}^\infty\frac{B_n(\kappa)}{p^2+m_n^2},\qquad
  B_n(\kappa)=\frac{(2n+1)^2\pi^3}{2(1-\kappa)K(\kappa)^3\sqrt\kappa}
  \frac{(-1)^ne^{-(n+1/2)\varphi(\kappa)}}{1-e^{-(2n+1)\varphi(\kappa)}},
\end{equation}
where
\begin{equation}
\varphi(\kappa)=\frac{K^*(\kappa)}{K(\kappa)}\pi,\qquad K^*(z)=K(1-z).
\end{equation}
Similarly, for the masses $m_n^2$ it is
\begin{equation}
\label{eq:spec}
m_n^2=m^2+\frac{Ng^2\mu^4}{m^2+\sqrt{m^4+2Ng^2\mu^4}}
\end{equation}
and $m^2$ is evaluated thorugh
the propagator~(\ref{Delta0mom}) as
$m^2=2Ng^2\Delta(0)$, yielding
\begin{equation}\label{m2}
m^2=-2iNg^2\int\frac{d^4k}{(2\pi)^4}\sum_{n=0}^\infty\Delta(k).
\end{equation}
For the scalar field, one has the solution given in Eq.\ (\ref{solG1}).
Similarly, for the scalar sector it is $G_2(x-y)=\Delta(x-y)$ provided in Eq.\ (\ref{m2}) one changes $2Ng^2\rightarrow 3\lambda$.

\section{False vacuum}\label{sec3}

\subsection{General technique}


We discuss the derivation of the effective potential, much in the same way is done in \cite{Coleman:1973jx}, using the exact solution for the theory provided in \cite{Frasca:2015yva}.
We emphasize that the general theory of the effective potential is built up on a generic classical solution, not necessarily a constant, as can be seen in \cite{Peskin:1995ev}. Indeed, it is quite common in literature to find studies using kink solutions or instantons \cite{Weinberg:2012pjx} that can be traced back to the work of Fubini \cite{Fubini:1976jm} and Coleman \cite{Coleman:1985rnk}. Therefore, it is interesting to analyze the behavior of tunneling for different sets of exact solutions.
We start from Ref.\cite{Frasca:2022kfy} and assume the following partition function for the scalar field
\be
Z[j]=\sum_{n=1}^\infty\left(\prod_{m=1}^n\int d^4x_m\right) G_n(x_1..x_n)
\prod_{p=1}^nj(x_p),
\ee
with $G_n$ being the $n$-point function. 
All the $n$-point correlation functions could be computed from it in principle. The effective action will be given by
\be
\phi_c(x)=\langle\phi(x)\rangle=\frac{\delta}{\delta j(x)}W[j]\,,
\ee
where
\be
W[j]=\ln Z[j].
\ee
By a Legendre transform we get $\Gamma$ as
\be
\Gamma[\phi_c]=W[j]-\int d^4xj(x)\phi_c(x).
\ee
Stated otherwise
\be
\Gamma[\phi_c]=\sum_{n=1}^\infty\left(\prod_{m=1}^n\int d^4x_m\right) \Gamma_n(x_1..x_n)
\prod_{p=1}^n\phi_c(x_p).
\ee
Therefore we get the following relations between $n$-point functions and the $n$-point vertices
\begin{widetext}
\bea
\label{eq:Gn}
G_2(x_1,x_2)&=&\Gamma_2^{-1}(x_1,x_2), \nonumber \\
G_3(x_1,x_2,x_3)&=&\int d^4x'_1d^4x'_2d^4x'_3\Gamma_3(x'_1,x'_2,x'_3)G_2(x_1,x'_1)G_2(x_2,x'_2)G_2(x'_3,x_3),
\eea
\end{widetext}
and so on. 
It should be noted that
\begin{widetext}
\be
\int dz\frac{\delta^2\Gamma[\phi_c]}{\delta\phi_c(x)\delta\phi_c(z)}
\left(\frac{\delta^2\Gamma[\phi_c]}{\delta\phi_c(z)\delta\phi_c(y)}\right)^{-1}
=\delta^4(x-y).
\ee
\end{widetext}
This gives for the effective potential
\begin{widetext}
\be
L^4V_{\rm eff}[\phi_c]=-\sum_n\frac{1}{n!}\int d^4x_1\ldots d^4x_n\Gamma_n(x_1,\ldots,x_n)
[\phi_c(x_1)-\phi_0(x_1)]\ldots[\phi_c(x_n)-\phi_0(x_n)],
\ee
\end{widetext}
where $L^{4}$ is the volume of the space-time, 
taken to be finite.

We have
\be
\left.\phi_c(x)\right|_{j=0}=\phi_0(x).
\ee
This is the 1P-function that in our case will depend on coordinates differently from the standard perturbative approach where is taken to be 0.
In momentum space we can write
\be
\label{eq:Vq}
V(\phi_c)=-\sum_n\frac{1}{n!}\left.\Gamma_n(q_1,\ldots,q_n)\right|_{q_i=0}(\phi_c-\phi_0)\ldots(\phi_c-\phi_0).
\ee

Our solutions of the scalar and Yang-Mills fields proved us the nP-correlation functions. Therefore, for an application of Eq.\ (\ref{eq:Vq}), we need to evaluate the $\Gamma_n$ vertex functions through the correlation functions. We perform this analysis in the following section.

\subsection{Non-perturbative results for false-vacuum decay}

We now apply the above technique to the theory (\ref{teor}) and write
\begin{widetext}
\bea
G_3(x-x_1,x-x_2)&=&-\int d^4x_3G_2(x-x_3)V'''[\phi_0(x_3)]G_2(x_1-x_3)G_2(x_2-x_3) \nonumber \\
&=&-\int d^4x_1d^4x_2d^4x_3G_2(x-x_3)V'''[\phi_0(x_3)]\delta^4(x-x_1)\delta^4(x-x_2)G_2(x-x_3)G_2(x-x_3).
\eea
In order to get $\Gamma_3$, the above equation and Eq.\ (\ref{eq:Gn}) must coincide. This can be obtained by
\be
\label{eq:Gamma3}
\Gamma_3(x_1,x_2,x_3)=-V'''[\phi_0(x_1)]\delta^4(x_1-x_2)\delta^4(x_1-x_3).
\ee
\end{widetext}
Then, by Eq.\ (\ref{eq:Vq}) till the third term gives the effective potential, that we call $U(\phi)$ to distinguish it from the general case, in the form
\be
\label{eq:FullU}
U(\phi_c)=V(\phi_c)-\frac{1}{2}C(R)Ng^2\Delta(0)\phi_c^2+\frac{1}{2}V_2(\phi_c-\phi_0)^2-\frac{1}{3!}|V_3|(\phi_c-\phi_0)^3,
\ee
where $C(R)$ is the Casimir index and $V_2$ and $V_3$ the coefficients we will give in a moment using the definition (\ref{eq:Vq}) and the just computed vertex functions $\Gamma_2$ and $\Gamma_3$.

The potential for the $\phi^4$ theory can be evaluated as follows
Using eqs.\ (\ref{eq:Gn}) and (\ref{Delta0mom}), one has
\be
V_2=\Gamma_2(0,0)=\frac{K(i)}{2\pi}\mathscr{A}\sqrt{2\lambda}\mu^2=\sqrt{2\lambda}\mu^2,
\ee
with
\be
\mathscr{A}=\left(\sum_{n=0}^\infty A_n\right)^{-1}=\frac{2\pi}{K(i)},\qquad
A_n=\frac{e^{-\left(n+\frac{1}{2}\right)\pi}}{1+e^{-(2n+1)\pi}}.
\ee
The next term is given by Eq.\ (\ref{eq:Gamma3}) in the form
\be
V_3=\Gamma_3(0,0,0)=-6\lambda\phi_0(0)=-3(2\lambda)^\frac{3}{4}\mu\operatorname{sn}(\theta,i).
\ee
The phase $\theta$ arises by integration of the equation of the 1P-function. We realize that quantum corrections break $\mathbb{Z}_2$ symmetry.

\begin{figure}[H]
\centering
\includegraphics[width=1\textwidth]{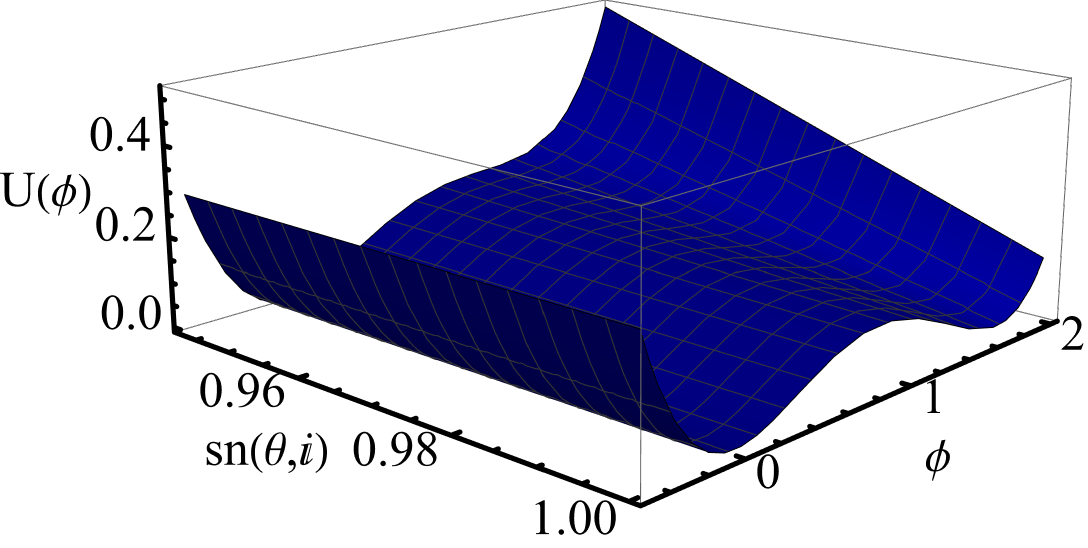}
\caption{\it Plot of the effective potential when the gauge theory is taken with a small coupling. Decay of the false vacuum can occur. This effect can be seen only for $\lambda>g^2/4\pi$.
\label{fig1}
}
\end{figure}

\begin{figure}[H]
\centering
\includegraphics[width=0.48\textwidth]{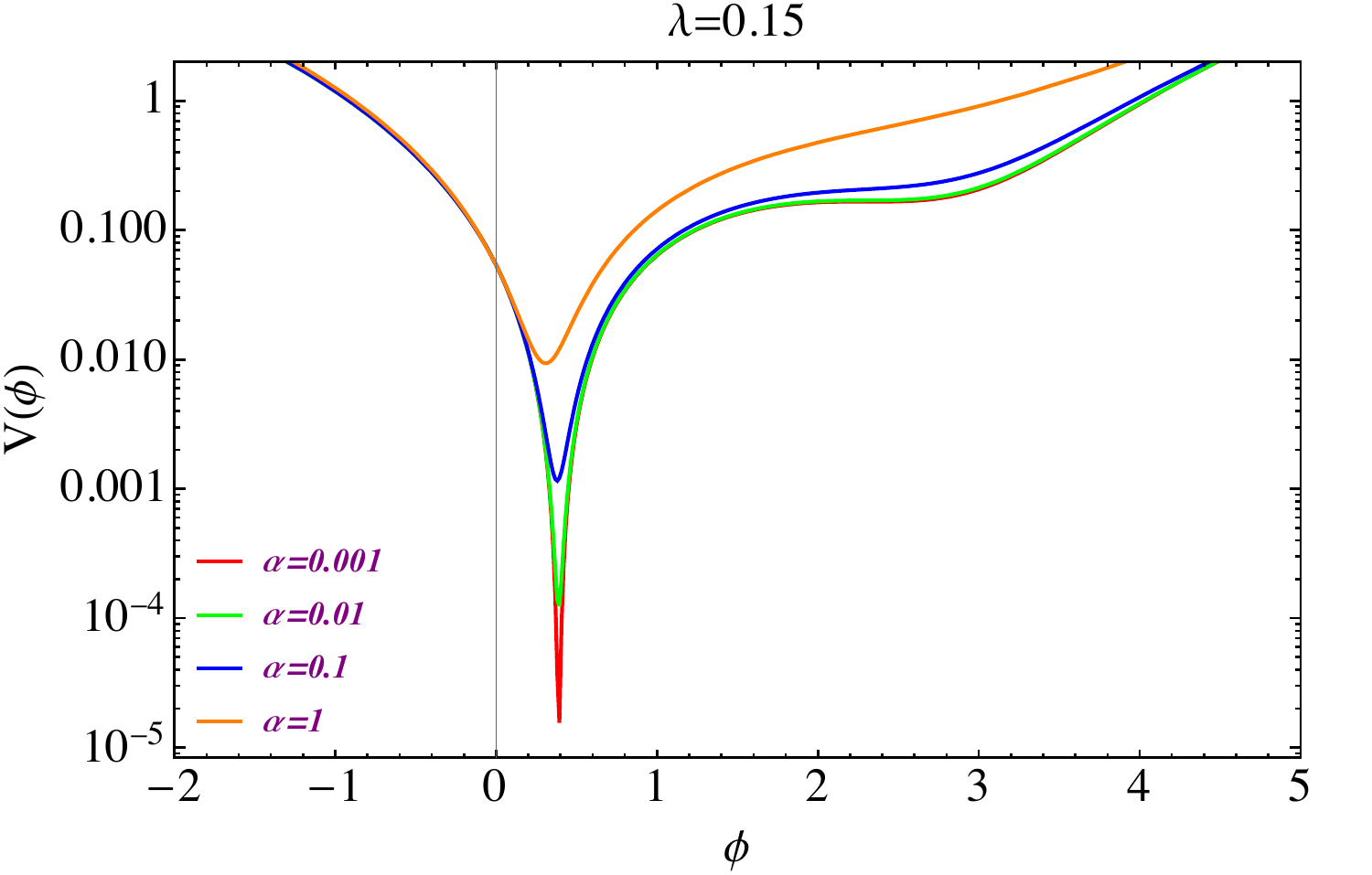}
\includegraphics[width=0.48\textwidth]{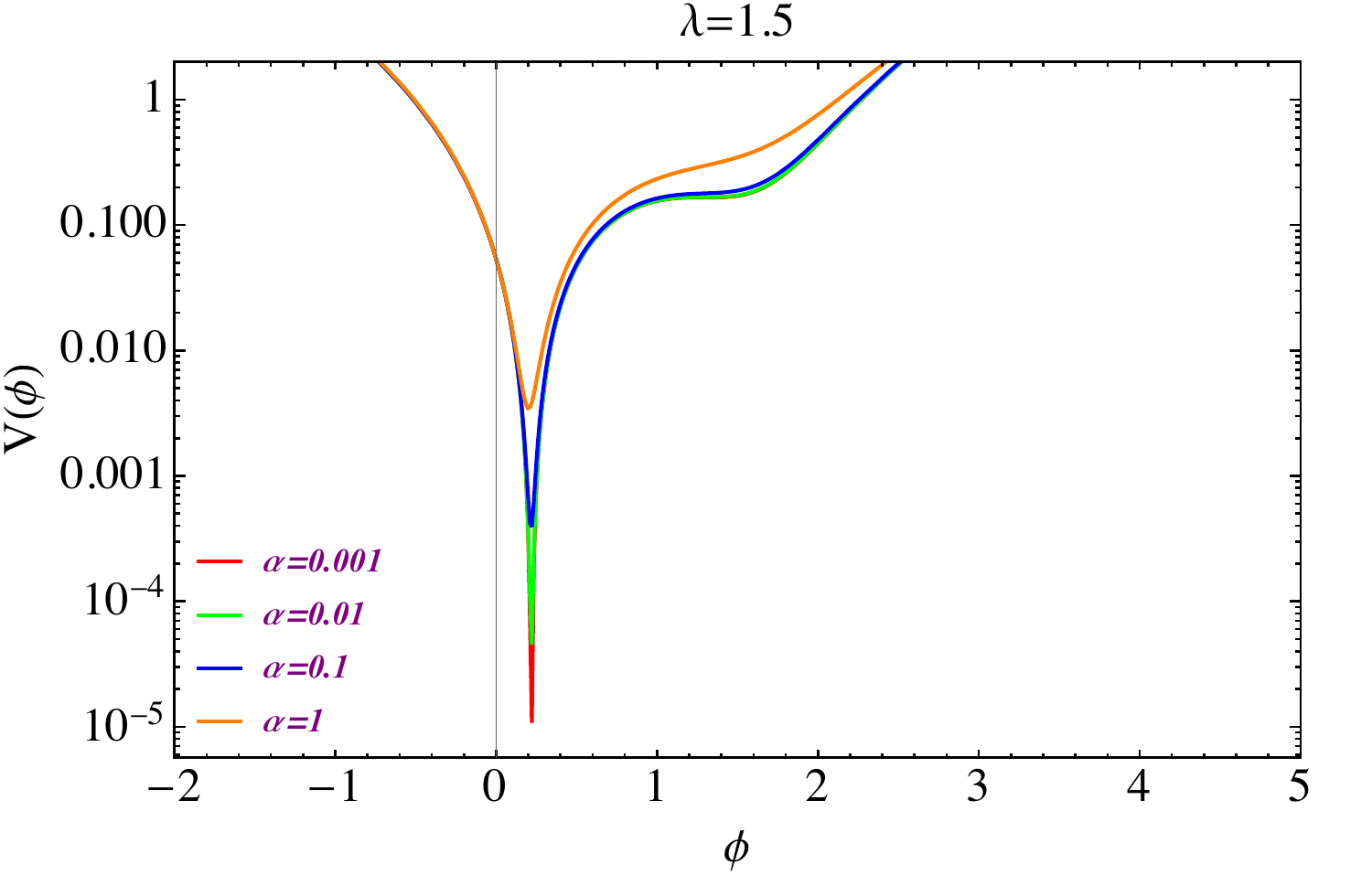}
\caption{\it Plot of the effective potential having fixed $\lambda$ (coloured curves) for different values of the gauge field coupling $\alpha=g^2/4\pi$. Here $\theta\approx 1.06966$ and $\alpha$ varies between $0.001$ and $1$ (color bars to recognize the single curves).}
\label{fig2}
\end{figure}

\begin{figure}[H]
\centering
\includegraphics[width=0.48\textwidth]{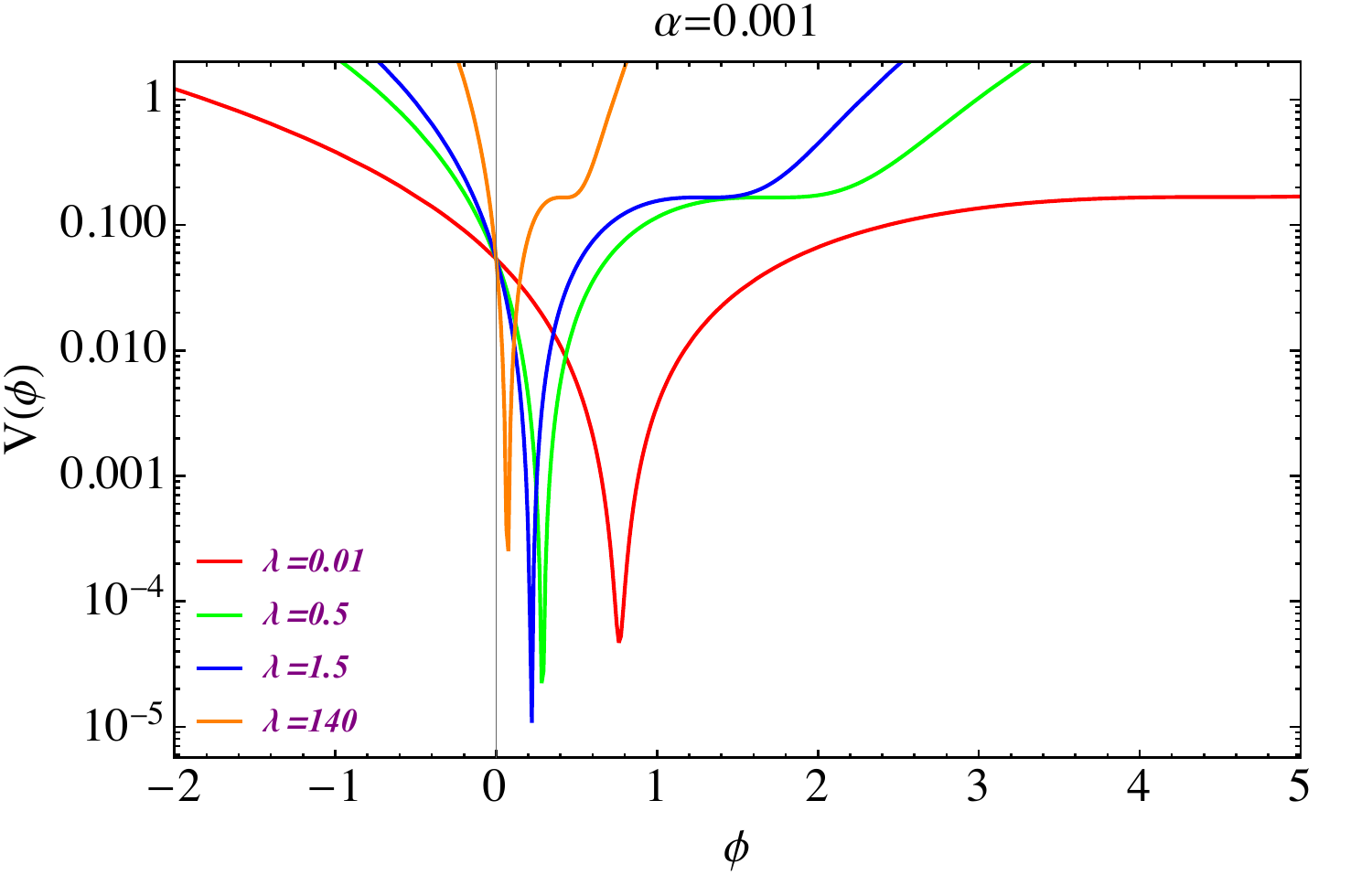}
\includegraphics[width=0.48\textwidth]{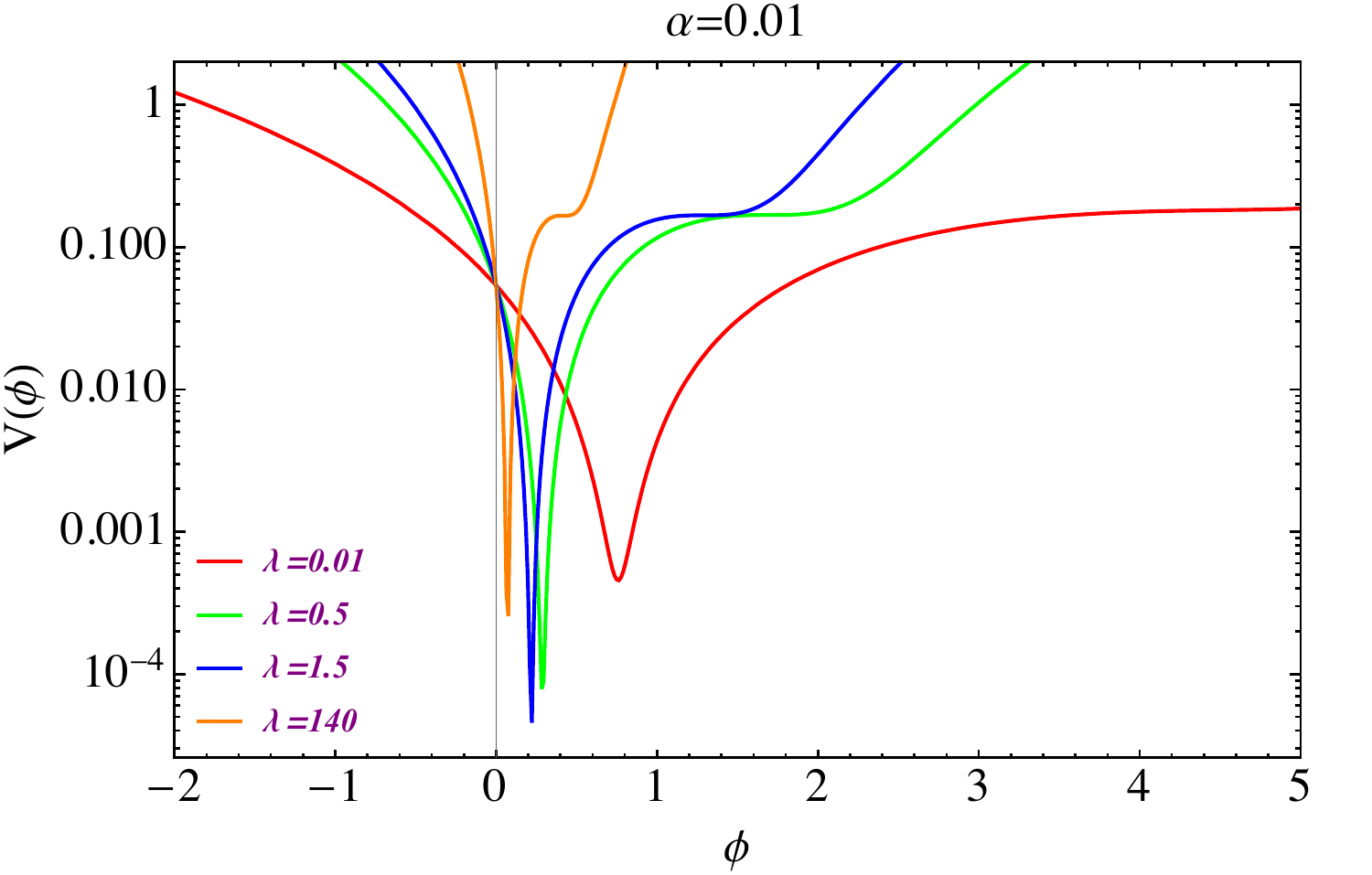}
\includegraphics[width=0.48\textwidth]{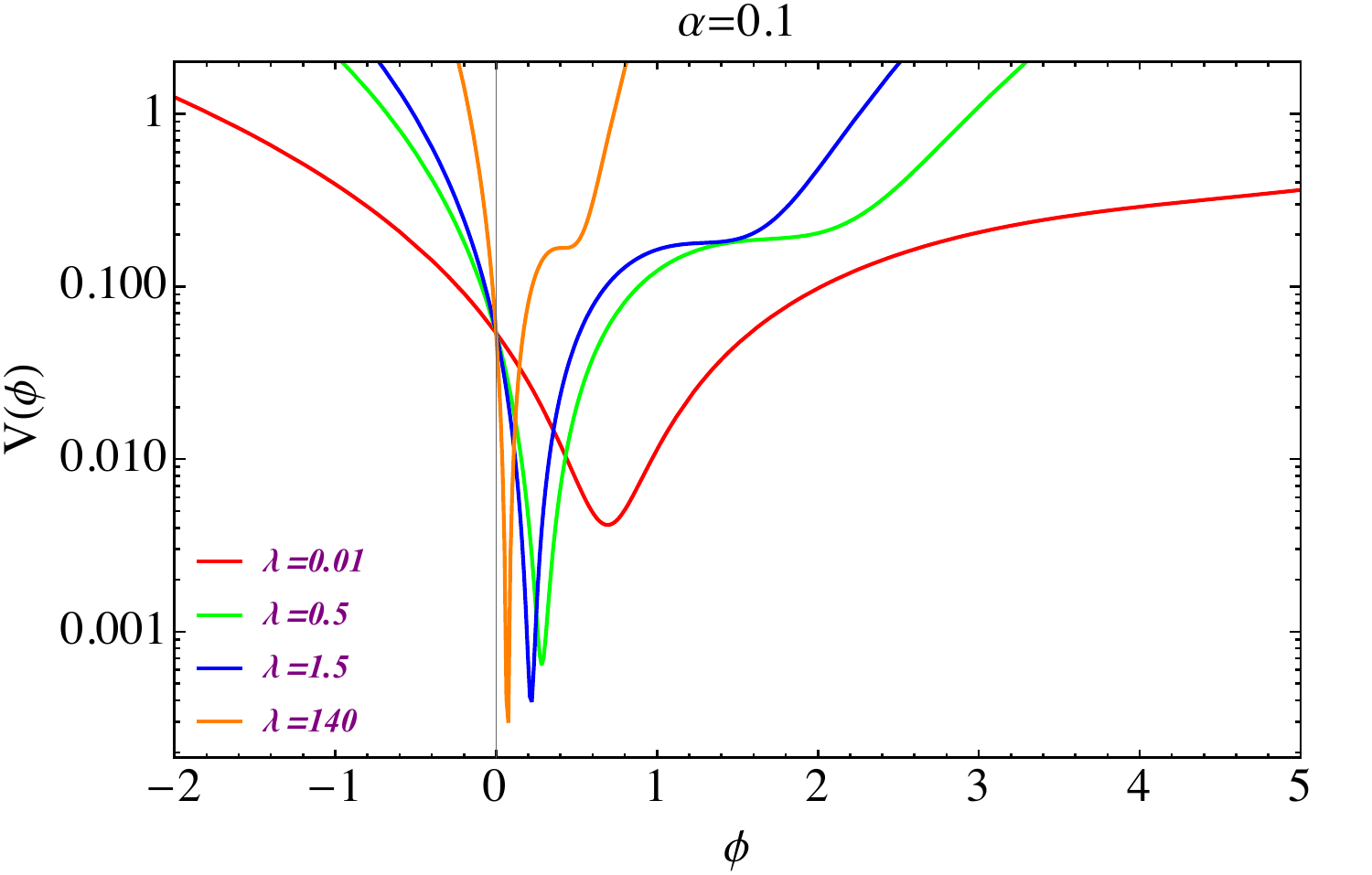}
\includegraphics[width=0.48\textwidth]{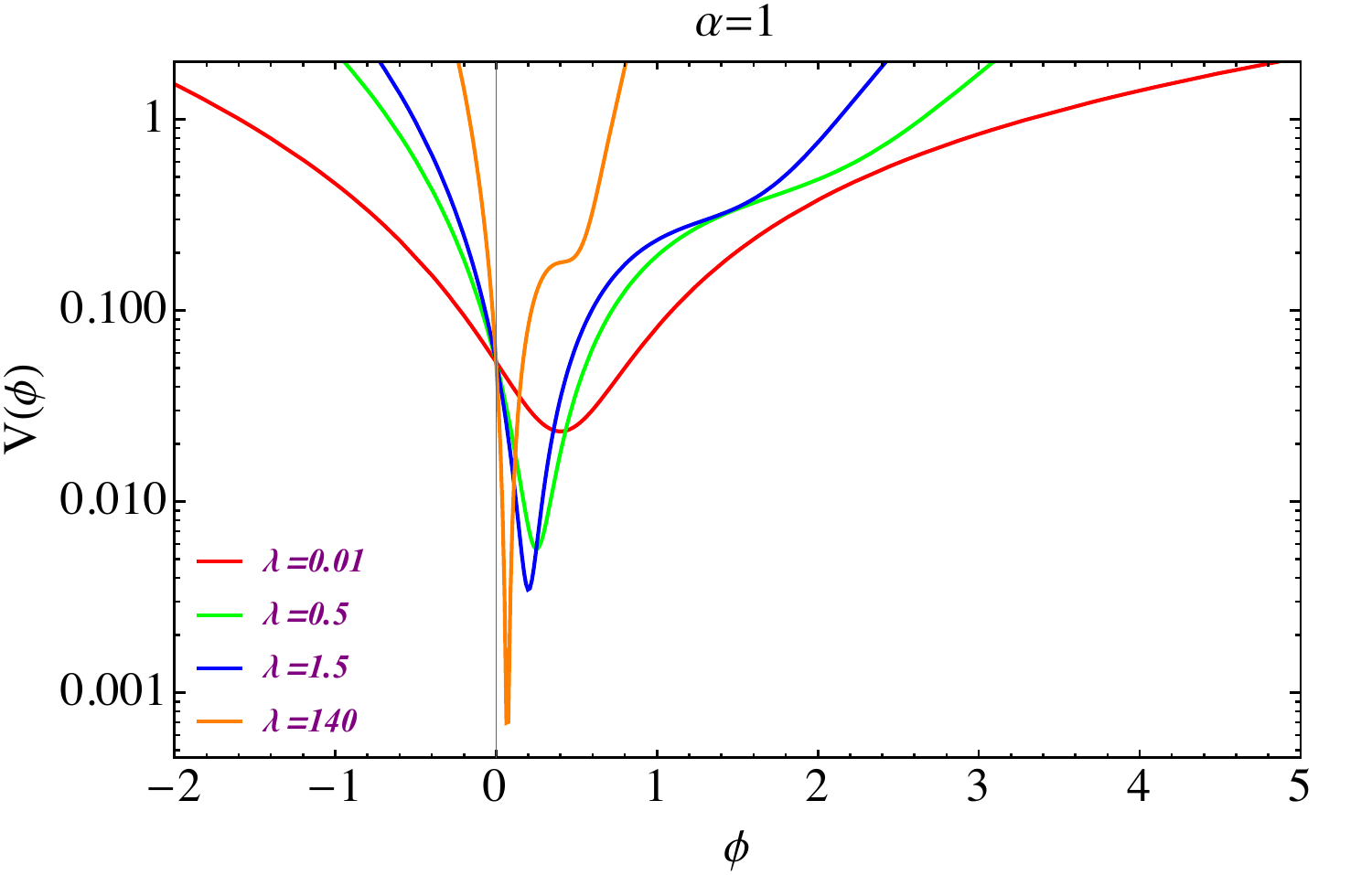}
\caption{\it Plot of the effective potential having fixed values of $\alpha$ (coloured curves) for different values of the scalar field self-coupling $\lambda$. Also here $\theta\approx 1.06966$.
}
\label{fig3}
\end{figure}

Finally, noticing that \cite{Frasca:2017slg}
\be 
\Delta(0)=-\frac{|\omega|}{2}\sqrt{\frac{Ng^2}{2}}\sigma_0,
\ee
where $|\omega|=0.03212775693\ldots$ and $\sigma_0$ is an integration constant of the gauge theory with the dimensions of a squared mass, we get the potential
\be
\label{eq:FullU_1}
U(\phi_c)=\frac{\lambda}{4}\phi_c^4+\frac{1}{2}\frac{|\omega|}{2}C(R)Ng^2\sigma\phi_c^2+\frac{1}{2}\sqrt{2\lambda}\mu^2(\phi_c-\phi_0)^2-\frac{1}{2}(2\lambda)^\frac{3}{4}\mu\operatorname{sn}(\theta,i)(\phi_c-\phi_0)^3,
\ee
where we have set as customary $\sigma = (Ng^2)^\frac{1}{2}\sigma_0$. This is proportional to the mass gap of the theory. The gauge field yields a mass to the scalar field that persists also around $\phi_c=\phi_0$. False vacuum decay appears just for the cubic term having the right sign and in the regime 
\be
\lambda>\alpha,
\ee
where $\alpha=\frac{g^2}{4\pi}$. 
The reason to stop to the fourth order for the potential arises from the assumption that the gravitational field, in a primordial universe, should introduce a proper mass scale that sets the UV-regime limit to a IR-regime where the scalar field becomes weak enough and a Higgs potential with a false vacuum could appear. 
%
A plot is given in Fig.\ \ref{fig1} clearly showing the appearance of a false vacuum at different values of the phase $\theta$ and the field $\phi$ having kept fixed both the couplings with the further condition $\lambda>\alpha$. 
Gauge theory at strong coupling, such as in a confining regime, can wash the effect out. This argument is fully consistent for the Standard Model. Similarly, false-vacuum decay is also washed out when the gauge coupling gets too high values with respect to the scalar field self-coupling, as can be appreciated in Fig.\ \ref{fig2}. One can see that some values of the background field can hinder false-vacuum decay by modifying the potential significantly.
Similarly, we have the plot in Fig.~\ref{fig3}.
From Fig.~\ref{fig2} and \ref{fig3} it is realized that increasing the gauge coupling removes the false vacuum and the decay is hindered and also non-existent. When the gauge coupling is smaller than the self-coupling of the scalar field, we get the inverse situation with the appearance of a false vacuum and a possible decay. This shows that a strongly coupled gauge field grants a stable vacuum.


In order to complete our analysis, we can evaluate the fluctuations entering into the decay rate from the standard formula \cite{Callan:1977pt}
\be
A=\left[\frac{\operatorname{det}'{-\partial^2+V''(\phi_0)}}{\operatorname{det}(-\partial^2+V''(\phi_+))}\right]^{-\frac{1}{2}},
\ee
where $A$ is dimensionless, the prime in $\operatorname{det}'$ means that we omit the zero mode, $\phi_0$ is the bounce and $\phi_+$ is the false vacuum. The spectrum of the operator is known: It is continuous and the eigenvalues are $3\mu^2\sqrt{\lambda/2}$ excluding the zero mode at $\mu=0$. Thus,
\be 
\operatorname{det}'[-\partial^2+V''(\phi_0)]= \prod_\mu\left(3\mu^2\sqrt{\frac{\lambda}{2}}\right)=\exp\left[\sum_\mu\ln\left(3\mu^2\sqrt{\frac{\lambda}{2}}\right)\right].
\ee
Let us introduce a cut-off $\Lambda$ that could be taken to be the Planck mass and promote the sum to an integral:
\be 
\prod_\mu\left(3\mu^2\sqrt{\frac{\lambda}{2}}\right)=\exp\left[\frac{1}{\Lambda}\int^\Lambda d\mu\ln\left(3\mu^2\frac{\lambda}{2}\right)\right]=\frac{3}{e^2}\sqrt{\frac{\lambda}{2}}\Lambda^2.
\ee
Similarly, we get by a textbook computation \cite{Peskin:1995ev} after renormalization
\be 
\operatorname{det}(-\partial^2+m^2)=\exp\left(\frac{Vm^4}{32\pi^2}\log(m^2/\Lambda^2)\right)
\ee
where $m^2=V''(\phi_+)$, $V$ the spacetime volume and $\Lambda$ is a cut-off. The final result is
\be 
A
=\left(\frac{e^2\sqrt{2}}{3\lambda}\right)^\frac{1}{4}\Lambda^{-1}\exp\left(\frac{Vm^4}{32\pi^2}\ln\frac{m^2}{\Lambda^2}\right).
\ee
This simple estimation shows that fluctuations increase as the self-coupling of the scalar field gets smaller.


\section{Conclusions}\label{sec4}

The most important consequence to extract from the above results is that there is a novel interplay between the self-coupling $\lambda$ of the scalar field and the coupling $g$ of the gauge field as evident from Fig.~\ref{fig2} and \ref{fig3}. If the latter prevails, the vacuum decay is hindered and possibly impeded. When the gauge coupling is smaller than the self-coupling of the scalar field, we observe the appearance of a false vacuum and its decay decay. This shows that a strongly coupled gauge field grants a stable vacuum.  This is an interesting result in view of the evolution of the universe. In fact, we can cautiously export these results to a scenario where the scalar and gauge fields live on a curved background, in which case the transition to the true vacuum is facilitated by gravity in Einstein's theory \cite{Calcagni:2022tls}. The picture that emerges is the following. At a very early stage when the temperature was higher, the scalar field coupling was higher than that of the strong interactions, possibly in a gluon-quark plasma state where the gauge coupling was very small. 
This scenario should be enforced by the asymptotic freedom regime that sets in at higher energies and, so also thermal corrections to the correlation functions should support it as seen from the QCD phase diagram in lattice computations \cite{Guenther:2020jwe}. 
%
This configuration favoured a false-vacuum decay. When the universe cooled down, the situation became inverted 
and the gauge coupling dominated,
converging the QCD phase to the confined one lowering temperature,
over the scalar one, thus preventing any further tunneling to lower vacua (which are absent in the quartic potential we considered here, but that could arise in a more complicated scenario). Therefore, the universe is very naturally driven to a stable state. 

The study of false-vacuum decay with gauge theory in the presence of gravity is in our future agenda to confirm this description of the early universe.

 \section*{Acknowledgement}

G.C.\ is supported by grant PID2020-118159GB-C41 funded by MCIN/AEI/10.13039/501100011033.

\end{document}